# Local electroexfoliation of graphene with a STM tip[*]

C. Rubio-Verdú[†], G. Sáenz-Arce[†,‡], J. Martinez-Asencio[†], D. C. Milan[†], M. Moaied[¥2], J. J. Palacios[¥2], M. J. Caturla[†], C. Untiedt[†]

**Abstract**

Graphite surfaces can be manipulated by several methods to create graphene structures of different shapes and sizes. Scanning tunneling microscopy (STM) can be used to create these structures either through mechanical contact between the tip and the surface or through electro-exfoliation. In the latter, the mechanisms involved in the process of exfoliation with an applied voltage are not fully understood. Here we show how a graphite surface can be locally exfoliated in a systematic manner by applying an electrostatic force with a STM tip at the edge of a terrace, forming triangular flakes several nanometers in length. We demonstrate, through experiments and simulations, how these flakes are created by a two-step process: first a voltage ramp must be applied at the edge of the terrace, and then the tip must be scanned perpendicularly to the edge. *Ab-initio* electrostatic calculations reveal that the presence of charges on the graphite surface weakens the interaction between layers allowing for exfoliation at voltages in the same range as those used experimentally. Molecular dynamics simulations show that a force applied locally on the edge of a step produces triangular flakes such as those observed under STM. Our results provide new insights towards surface modification that can be extended to other layered materials.

---

[*] Electronic supplementary information (ESI) available

[†] Departamento de Física Aplicada. Facultad de Ciencias, Universidad de Alicante, E-03690, Alicante, Spain, e-mail: mj.caturla@ua.es

[‡] Laboratorio de Materiales Industriales, Departamento de Física, Universidad Nacional, 86-3000 Heredia, Costa Rica

[¥]Departamento de Física de la Materia Condensada, Instituto de Ciencia de Materiales Nicolás Cabrera (INC), and Condensed Matter Physics Center (IFIMAC), Universidad Autónoma de Madrid, Cantoblanco, 28049 Madrid, Spain

**INTRODUCTION**

Its low reactivity and almost perfect flatness makes highly ordered pyrolytic graphite (HOPG) surfaces ideal to be used as substrates for scanning tunneling microscopy (STM) studies even at ambient conditions[1,2], easily achieving atomic resolution. Moreover, a STM can also be used to modify the surface of graphite in different ways[3-5]. Albrecht et al. demonstrated the possibility of creating holes[3] on a graphite surface by simply applying positive voltage pulses with the STM tip. Kondo et al.[4] showed that while positive voltage pulses produce etching, negative pulses give rise to metallic deposits. The formation of holes was explained as a result of the sublimation of carbon atoms induced by tunneling electrons when this process was done in UHV conditions, and to chemical reactions on the surface for ambient conditions. Later, Hiura[5] demonstrated that also the scanning speed affected the results showing a relationship between the threshold voltage to produce a hole and the scanning speed. They attributed this dependence to the kinetics of the oxidation process. In all the cases above, the procedure in which the graphite surface was modified involved either the mechanical interaction of the STM with the surface, thermal sublimation, or an induced chemical reaction.

But it was D. Eigler in his pioneer work[6] who showed that an electrical field can also be used to attract and displace atoms over a surface. This effect has also been used to displace larger objects[7] including graphite flakes[8], and on macroscopic HOPG substrates where high voltage differences between two electrodes have been used to exfoliate a few layers of graphite[9-11]. The mechanisms involved in such process are not fully understood. Here we show how applying a voltage ramp at the edge of a terrace in graphite and then scanning perpendicularly to the edge, triangular flakes of graphene appear on the surface that can be up to several nanometers long. *Ab initio* calculations and molecular dynamics simulations shed some light on how these structures are formed.

## METHODS

### Experimental set up

For our experiments we have used a homemade STM. The STM is controlled using Nanotec's "Dulcinea" control unit electronics and all the images have been analyzed using the Wsm[12] software. For the samples we have used HOPG (Goodfellow, 10x10x2mm), which is cleaved by mechanical exfoliation using the Scotch tape method. The tips were of an Ir-Pt alloy scissor-cut prior to the experiments. Experiments were performed both in air and under vacuum conditions (P ~ $10^{-5}$ mbar), using a turbomolecular pump. Most of the experiments have been performed at room temperature. Similar results were obtained at 77K.

### Electrostatic calculations

Our calculations are based on density functional theory (DFT)[13, 14] as implemented in the SIESTA code[15, 16]. We are mostly interested here in multilayer graphene and graphite where dispersion (van der Waals) forces due to long-range electron correlation effects play a key role in the binding of the graphene layers. Therefore, we use the exchange and correlation nonlocal van der Waals density functional (vdW-DF) of Dion et al.[17] as implemented by Román-Pérez and Soler[18].

To describe the interaction between the valence and core electrons we used norm-conserved Troullier-Martins pseudopotentials. The cutoff radii were 1.56 Å for both the s and p components in C, and 1.25 Å for the s component in H[19]. To expand the wavefunctions of the valence electrons, a double-Z plus polarization (DZP) basis set was used[20]. We experimented with a variety of LCAO basis sets and found that, for both graphene and graphite, the DZP produced high-quality results.

The plane-wave cutoff energy for the wavefunctions was set to 500 Ryd. For the Brillouin zone sampling we use 4x4 Monkhorst-Pack k-mesh for the 12x12 multilayer graphene supercells. We have also checked that the results are well converged with respect to the real space grid. Regarding the atomic structure, the atoms are allowed to relax down to force tolerance of 0.005 eV/A.

For charged systems all supercells are separated by a vacuum space of at least 99 Å so that the interaction between charged graphene layers and their periodic images can be safely ignored as the surface layer is separated from the multilayer system[21, 22]. We have considered positive charging, i.e., electron depletion (Q > 0), expressed in units of electron charge (e) per surface atom in the figures.

**Molecular dynamics calculations**

The molecular dynamics simulation package LAMMPS[23] was used for the simulations of the lifting and tearing of graphite layers with the AIREBO interatomic potential[24]. The cutoff used for the interatomic potential was 12 Angstroms to reproduce the equilibrium distance between graphite layers. The simulation set up consists of eight graphite layers of a 10 nm x 10 nm surface, with the last layer being a step of one monolayer, and a total of 28800 atoms (see figure of set up in supplementary material ESI S1). A force is applied right at the edge of the step, in the middle of the simulation cell, within a semi-sphere of 1 nm diameter. This force mimics the force applied by the STM tip. Values of the force between 1 nN/atom and 6 nN/atom have been used in the simulations. Free surfaces are considered in the z direction and periodic boundary conditions are applied in the x and y directions. To avoid pulling completely the last monolayer, atoms at the border of the simulation cell are kept fixed. Two types of steps are considered, one ending on a zigzag configuration and one ending on an armchair configuration. Two different types of simulations are performed. In one case, an upward force is applied for 1000 steps (of 1 fs per time step), then the value of the force is reduced to 0.6 – 0.8 nN per atom for another 500 steps and finally the system is relaxed for 2000 steps. This simulates the voltage ramp used experimentally. In a second set of simulations the force is applied parallel to the surface and perpendicular to the step. The force is applied until the pulled layer reaches the limit of the simulation cell. These second type of simulations mimic the scanning of the STM tip after the voltage ramp.

**RESULTS AND DISCUSSION**

For these studies we place the STM tip at the edge of a graphite terrace as the one shown in Figure 1a. The mechanism we use is conceptually very simple and consists of gradually increasing the applied voltage between the STM tip and the graphite surface, thus increasing their electrostatic attractive force. When doing this while keeping a low and constant tunneling current (about 0.1 nA in our case) to avoid any contact between tip and sample, we can observe that at a certain applied voltage the STM tip shows a sudden retraction, meaning that the graphite surface is being lifted. Afterwards an image of the surface is taken by scanning the tip, showing a triangular flake folded on top of the original graphite terrace as shown in Figure 1b. Other examples of nanostructures fabricated using this method are given in the supplementary material (ESI Figures S2 – S4).

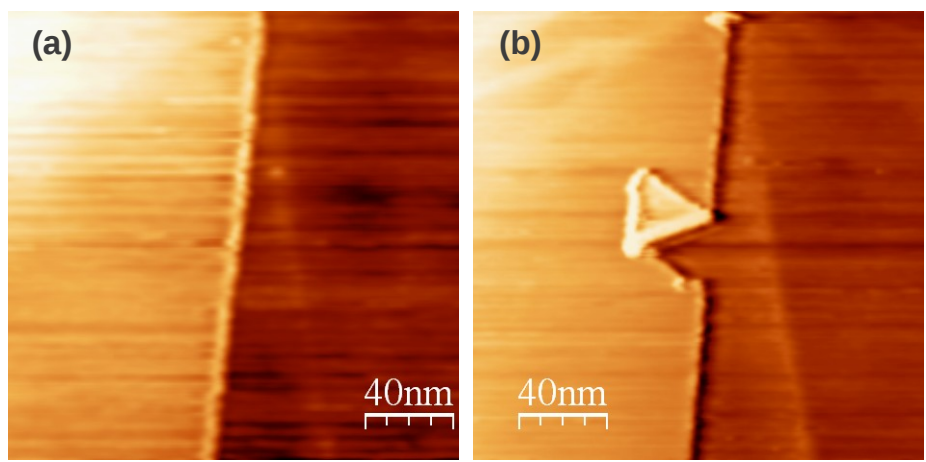

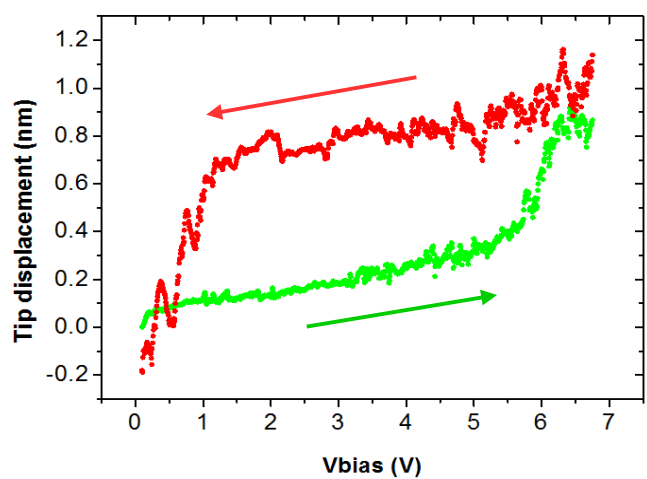

**Figure 1:** An example of the initial surface showing a terrace (a) and the same surface after the electro-exfoliation process, that is, applying a voltage ramp and obtaining an image of the surface by scanning the tip (b). The formation of a triangular flake, folded over the original terrace is clearly shown (c) Recorded values of the applied voltage and the distance between the tip and the surface (height) for a particular experiment. Green curve are values for increasing voltage bias while red curve are for decreasing voltages.

The whole process can be characterized by recording the movements of the STM tip as depicted in Figure 1c. The green curve starts at about 0.1 V that is the applied voltage used for imaging the graphite surface with our STM. As the voltage is increased, and in order to keep the current constant, a slight continuous retraction of the STM tip is observed. Note that during this whole process the tip is kept on the same location over the edge of the graphite terrace. At about 3 V the tip has been retracted approximately 0.1 nm, less than one graphite terrace height. This seems to be a common feature of all the cases analyzed in detail. From 3 V onwards the behavior changes from one case to another. In the particular case presented in Figure 1c the tip retracts ~ 1 nm for a total applied voltage of 7 V. At ~ 3 V the retraction of the tip accelerates meaning that the last graphite layer is being lifted until the tip is retracted other 0.25 nm at about 6 V. This is the point when a sudden abrupt retraction, of about 0.5 nm in this case, is observed meaning that the graphite layer has been torn and lifted up. Note that, according to Kondo[4], the minimum voltage to produce a hole on a flat graphite surface under UHV is 8.5 V. In our experiments, we observe significant changes in the position of the tip for voltages between 4 – 6 V, lower than the value obtained by Kondo, most likely since the tip is located at the step edge, and not in the middle of a flat surface.

If the applied voltage is now decreased (red curve in Figure 1c), we observe almost no further displacement of the tip, what can be understood if the graphene layer has been indeed lifted irreversibly and the electrostatic force is still strong enough to keep this graphene layer lifted. At approximately 1 V, in this case, the layer falls back onto the surface. In other cases this lift-off of the graphene layer can also occur stepwise (see supplementary material ESI S4-S5). The exact details of this process depend on the tip used since its shape will determine the electrical field, which will remain constant for each tip.

It is important to note that the displacement of the STM tip due to the applied voltage in some cases is on the order of 1 nm, as shown in Figure 1c, while in other cases the tip can retract up to 4 nm (see ESI Figure S3 from the supplementary material). Interestingly, the flakes observed after scanning the surface to make an image are on the order of 20 nm, and in fact flakes up to 40 nm have been obtained.

This implies that in most cases both the voltage ramp and the subsequent scanning of the surface are contributing to the formation of these structures. In order to determine the effect of the scanning-tip, we have applied voltage ramps and studied the structures observed when scanning in a parallel or in a perpendicular direction to the graphite terrace edges. Large flakes were only obtained when the scanning was done perpendicular to the edge, whereas when the scanning is performed parallel to the edge these were not observed. Instead, some small holes or local irregularities appear on the surface. Nevertheless, we have to point out that in all cases the voltage ramp was necessary to obtain any surface modifications.

In this way, for most cases, we can understand the formation of these structures as a two-step process. Firstly, the voltage ramp decouples the upper graphite layer and starts a tear or forms a defect on it. Secondly, the scanning tip interacts with this first tear-seed, which propagates with the help of the scanning tip. This two-step process is clearly seen in the sequence of images presented in Figure 2. Figure 2 (a) shows the terrace before applying the voltage ramp. The first image taken after the voltage ramp shows the formation of a defect at the edge (Figure 2 (b)) while the tip has only retracted ~ 1 nm starting at a voltage of ~ 4 V (see ESI Figure S5 in the supplementary material). This implies that the layer has been slightly lifted and detached from the graphite substrate and a defect has been created. Further scanning of the surface produces a large flake ( ~ 20 nm in length) where the defect was first produced during the voltage ramp (Figure 2(c)). Another case is shown in the supplementary material where the tip retracts 4 nm and the flake formed is 11 nm (ESI Figure S4). Figure 2(d) shows a flake obtained from the molecular dynamics simulations performed as explained below. Note the resemblance (albeit at a different scale) between the experimental and simulated flakes.

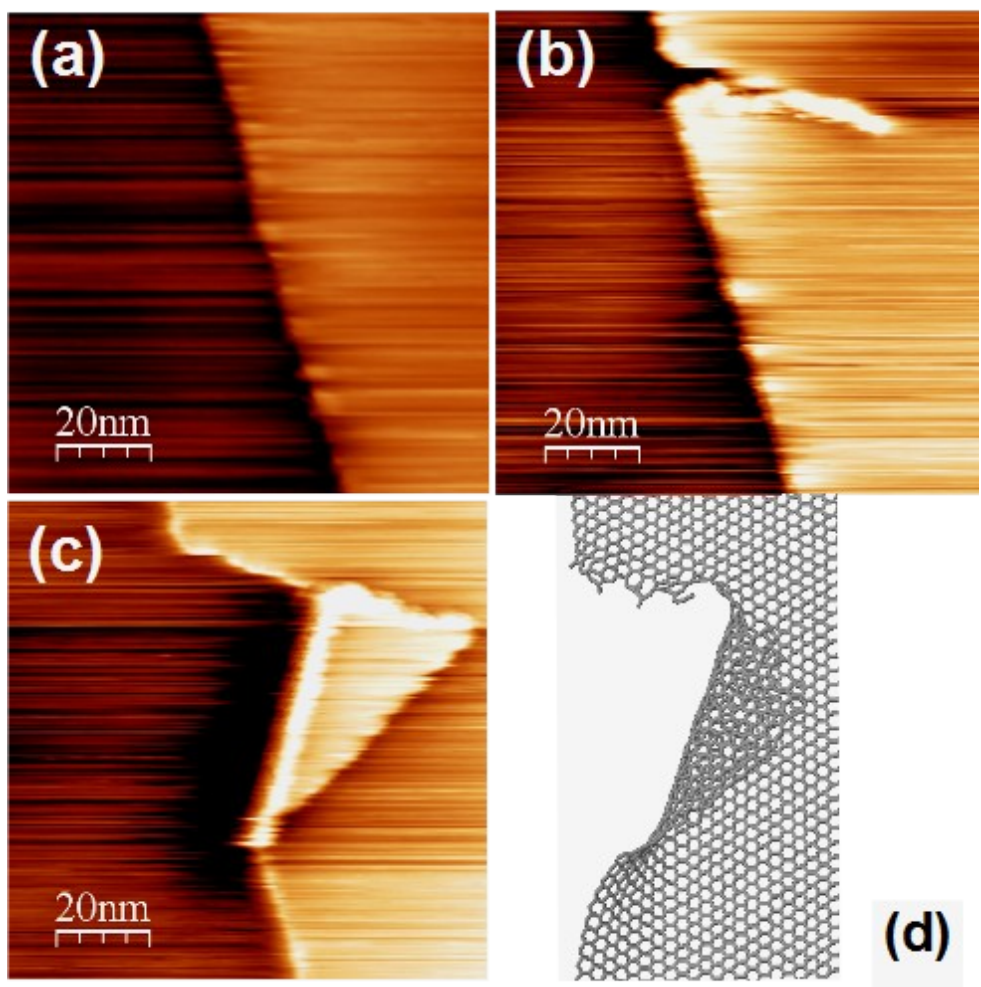

**Figure 2:** The three panels show a terrace in the process of being folded. (a) Terrace before applying a voltage ramp. (b) Defects are formed by the voltage ramp, which also detaches partially the flake from the substrate. Finally the movement of the STM tip, while scanning, folds the last graphite layer forming a triangular flake (c). Figure (d) shows one example of a flake obtained from molecular dynamics simulations.

The experimental observations can be rationalized in the following manner. A voltage applied between the STM tip and the surface induces charges on the surface of graphite (and on the tip) and the electrostatic attraction between these charges is ultimately responsible for the detachment of the surface layer (see a schematic representation in Figure 3a). One should also consider the fact that the accumulation of charge on the surface modifies the binding forces between the surface layer and bulk graphite. To model this at a fundamental level we have performed density functional theory (DFT) calculations, including van der Waals interactions, as explained above.

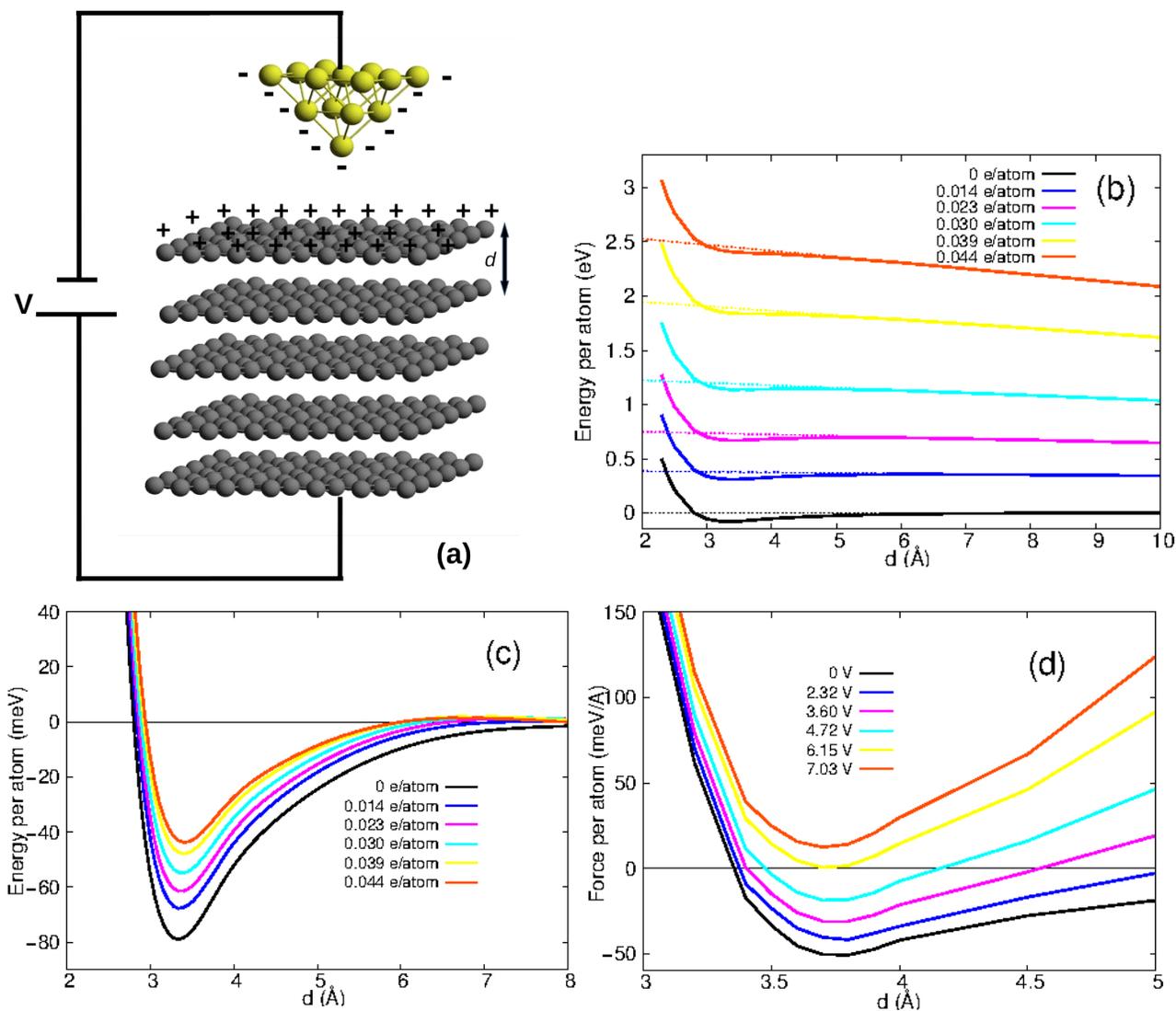

**Figure 3:** (a) Schematic picture of a STM tip on a graphite surface with an applied bias voltage inducing surface charges. (b) Energy per atom as a function of the distance between the surface layer and the rest of the multilayer system. Different colors correspond to different (positive) charges on the surface and the bottom layers. The dashed straight lines are fits to the long-range behavior. (c) Binding energy curves once the long-range behavior has been subtracted. (d) Force per atom as a function of the bias voltage applied between the multilayer system and a flat metallic tip.

The *ab-initio* modeling of the electrostatic exfoliation phenomenon requires a non-equilibrium DFT calculation since this effect results from the application of a bias voltage between tip and substrate. Alternatively we make use of standard equilibrium DFT calculations and simple electrostatic models as we explain in what follows. We first compute the (binding) energy of the top graphene layer as a function of the distance $d$ to a multilayer structure representing graphite, which is composed of 4 graphene layers, as the one shown in Figure 3(a). The calculated binding energy curve is given in Figure 3(b) for a neutral system (black line) and for a positively charged system where electrons have been removed (color lines). The extra (positive) charges logically appear at both surfaces of the multilayer system, on the fixed bottom layer and on the movable surface one. The long-range electrostatic repulsion between the charged surface layers at opposite sides is evident from the linear decay of the binding energy vs. distance in these cases. We now remove from all these curves this long-range contribution from the binding energy that is expected to depend linearly with the distance, $A+Bd$. The value of the constants $A$ and $B$ has been obtained by fitting the energy curves for distances longer that 0.6 nm, where van der Waals attraction forces begin to fade away (see black line for the neutral case), but never larger than 2.0 nm where the electrostatic interaction between supercells (placed at a relative distance of 10 nm) appreciably changes the linear behavior[21]. The resulting fitting is shown in Figure 3(b) by straight dashed lines and the result of subtracting this long-range contribution to the binding energy is shown in Figure 3(c). These curves represent now the binding energy of *charged* graphene layers to bulk graphite free of unwanted electrostatic interactions with the bottom layer. As can be seen in Figure 3(c) the new binding energy decreases as the surface charge increases. Finally, we include the effect of the STM tip, which is actually the one responsible for the charging of the surface layers. We assume a flat metal plate as a model for the STM tip located at a distance of $D_{tip}$ = 0.7 nm from the graphite surface. As a result of the applied bias voltage, which is chosen as to generate the surface charge that we have considered in the previous calculations, an attractive force appears between the oppositely charged plate and the surface layer as presented in Figure 3(d). The critical voltage at which the force becomes always positive (and therefore the surface layer necessarily detaches from the rest of the layers) lies between 6 and 7 V, which is within the range of the experimental values described above. This critical voltage obviously depends on the chosen distance between tip and surface as well as on the exact shape of the tip. More realistic models for the tip such as round or sharpened forms, acting more locally, may decrease these critical values giving closer values to those measured experimentally. Nevertheless, these simulations clearly show that the voltage ramp applied in the experiments is able to detach a graphene layer.

The dynamic processes that involve the tearing of the graphene flake and the actual exfoliation even have been simulated using molecular dynamics with empirical potentials. Firstly, an upward force is applied at the edge of a step ending either on a zigzag or an armchair configuration. This simulation mimics the force induced by the voltage ramp in the STM experiment. We have also performed calculations where the force is applied parallel to the surface and perpendicular to the terrace, to reproduce the conditions during the scanning of the surface by the STM. Details of the simulation set up are given in the supplementary material ESI S1.

In the process of lifting and breaking a layer of graphite we have seen that if the force is large enough, the graphite layer is torn close to the location of the applied force, eventually folding over itself. In most of the cases studied, tearing occurs not only close to where the force is applied, but also at the edges of the simulation box, where the layers are fixed. If the force is slightly lower, tearing occurs only at the edges, similarly to the simulations of Sen et al. [25]. If the force is lowered even further, the layer is only slightly lifted. An example of these simulations is included in the supplementary material ESI Figure S6 and video S1. These simulations reveal that it is possible to break and form triangular flakes by lifting the graphene layer vertically. In this case the layer is lifted ~ 4 nm. However, in most of the experiments, the tip only retracts on the order of 1 nm during the voltage ramp. If in the simulations the layer is allowed to relax after being lifted only about 1nm, small defects can be observed at the edge of the terrace but not nanometer size triangular flakes, consistent with the two-step process described above.

The second set of simulations study the effect of the scanning of the STM tip on the surface with a zigzag or armchair edge configuration, that is, the second step in the experiments shown above. The simulation set up is the same as in the case of lifting, but now the force is applied parallel to the surface and perpendicular to the step. Calculations have been performed with the presence of a small defect at the edge, which would be the result of lifting the layer due to the voltage ramp as explained above. Figure 4 shows the result of an armchair type of edge with a defect and a force applied next to the defect of 1.9 nN/atom in a radius of 0.5 nm. In this case the size of the simulation box is 20 nm x 20 nm with 5 graphite layers. A movie of this simulation is included in the video S2. First the layer is pulled to the side without breaking (figure 4(a)), and then it starts tearing at the location of the defect, and forming a triangular flake (figure 4(b)). And it continues breaking forming two flakes (figure 4(c)). One of them (left side of the image) starts breaking from the fixed sides, which is a consequence of the simulation size, losing its triangular shape, while the other one remains as a triangle (right hand side of the image). The calculation is stopped when the flake reaches the limits of the simulation cell. In figure 4 (d-f) a side view of the simulation box is presented for the same cases as figures 4 (a-c). These images show that in this case the height of the layer is on the order of 1 to 2 nm, at most.

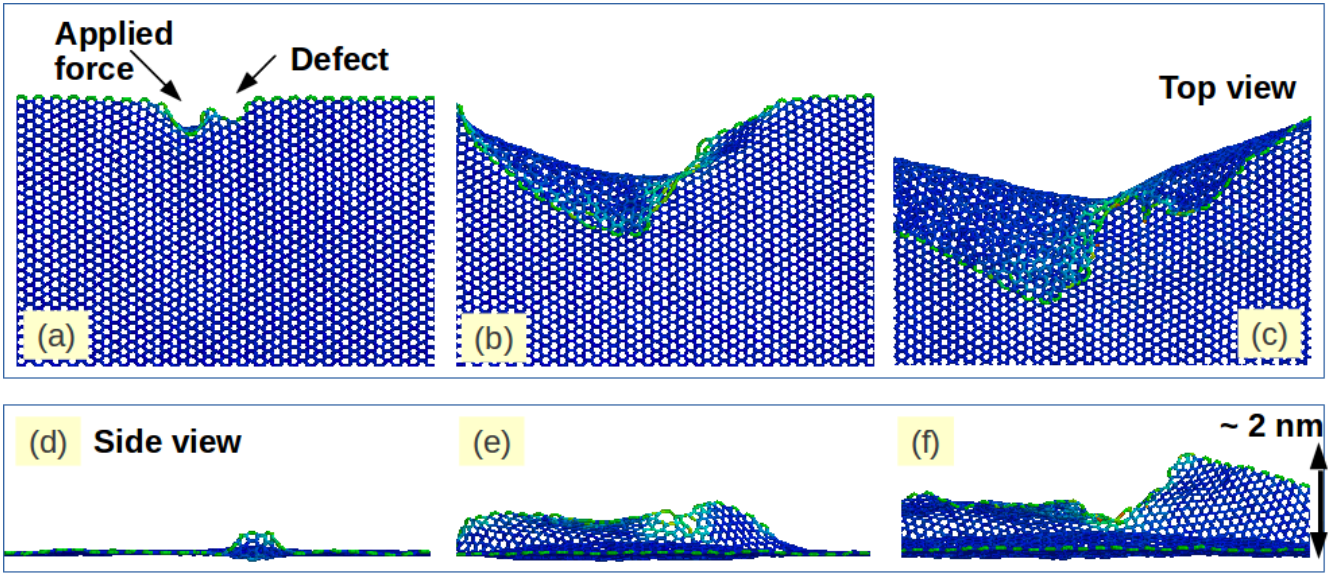

**Figure 4:** Formation of flakes on an armchair edge configuration applying a force of 1.9 nN/atom in a 0.5 nm radius parallel to the surface and perpendicular to the edge and next to a defect formed by 10 vacancies at the edge. Only the top layer of the graphite system is shown. First the layer is moved back without breaking (a), then it starts breaking by the location of the defect (b) forming two flakes that continue folding. One of the flakes breaks from the side, where the atoms are fixed (c). Side views of the same snapshots are shown in figures (d, e and f). Note that in this case the layer is lifted up 2 nm at most. Colors represent potential energy per atom.

Simulations also show that, consistently, a higher force must be applied to break a step ending on an armchair configuration than one on zigzag. For example, for an applied upward force of 2.2 nN per atom in a 0.5 nm radius a zigzag configuration breaks, while a force of 2.7 nN per atom must be applied in the same area when the step ends on an armchair configuration. We should mention that, on one hand these values decrease when the radius is increased and, on the other hand, the actual values of the force to break the layer change between different simulations where the only difference is the initial distribution of velocities. Nevertheless, the range of values is always lower for a zigzag configuration than for an armchair one.

In our experiments, another interesting feature observed is that all the folded graphene layers obtained show a triangular shape. In order to characterize and understand the origin of such shape prior to the folding process we have performed atomic resolution images of the surface (inset on Figure 5(a)), which is extrapolated to the whole graphene sheet (Figure 5(b)) to get the edge atomic structure of the flakes as shown in Figures 5(c-d). We noticed that all the terraces with straight termination showed either a zigzag or an armchair configuration. For all the cases examined, we saw that the tearing process occurred always through the zigzag direction while the graphene layer was bent through this same direction independently of the starting edge configuration, as Figures 5(c-d) show. As a consequence, two possible triangular shapes are observed depending on the starting edge structure, an equilateral when starting from a zigzag or a 60-30-90 triangle when starting from an armchair.

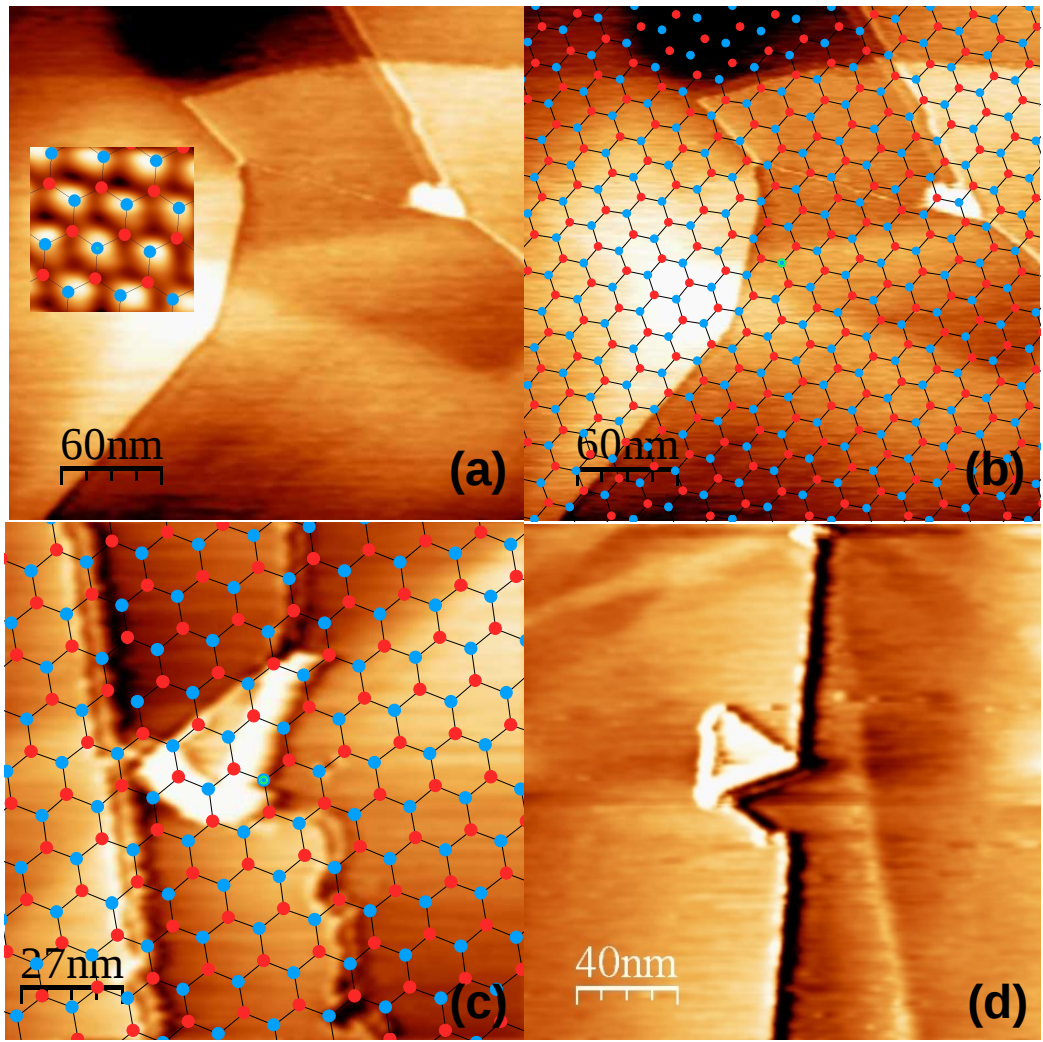

**Figure 5:** From the atomic resolution image (inset in a) the edge structure of the graphite terraces can be characterized as the lattice is extrapolated to the whole image (b) . In this way we can characterize the flake geometry when the original terrace had an armchair edge atomic structure (c) or the flake geometry when the original terrace had a zigzag edge configuration (d).

Other groups have observed similar triangular shapes when graphene is scraped off graphite on macroscopic graphene sheets[25] or through atomic force microscopy[26] . The formation of these triangles has been explained as the interplay between three energies: elasticity, adhesive energy and fracture energy[25] . Tears in suspended graphene have been studied both theoretically and experimentally by Kim et al[27] . They observe that torn edges are generally straight along zigzag or armchair directions, with armchair ones happening twice as often as zigzag, and with occasional changes in direction of 30º. This result differs from the observations in our experiments, where the preferential direction for tearing the graphene layer is the zigzag direction. In our case the possibility of performing atomic resolution images has allowed us to unambiguously determine both tearing and folding direction.

Looking in detail at the triangular shapes formed in the simulations we observe that breaking occurs along the zigzag direction independently of the edge finishing in an armchair or in a zigzag configuration (see figure 6(a)). The bending of the layer is also preferentially along the zigzag direction in the case where the force is applied parallel to the surface (see figures 6(a-b)). When the force is applied upwards there are some cases where the bending occurs along the armchair direction (see Figure 6(c)). These calculations support the explanation given above for the formation of these triangular flakes with a STM.

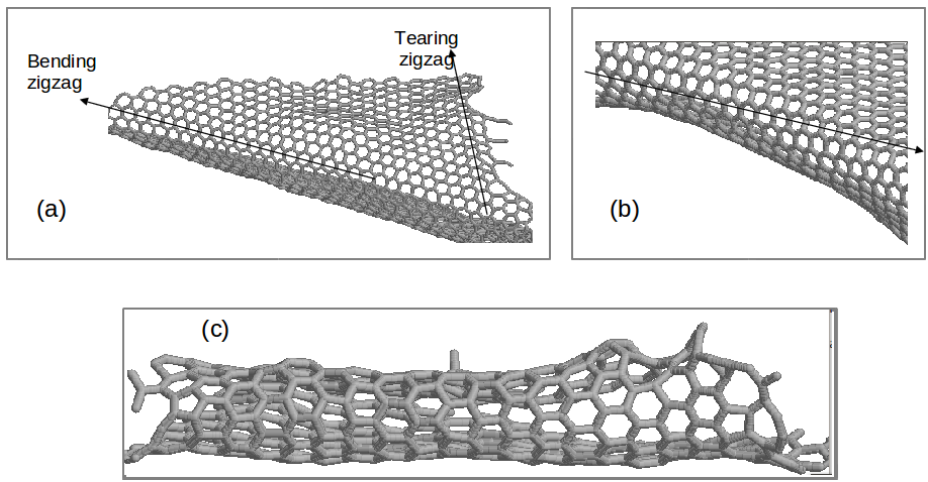

**Figure 6:** Examples of flakes from simulations (a) armchair edge configuration and parallel force of 1.9 nN/atom in a 0.5 nm radius (b) zigzag edge configuration and parallel force of 1.6 nN/atom in a 0.5 nm radius and (c) zigzag edge configuration and an upward force of 2.2 nN/atom in a 0.5 nm radius. Cases (a and b) show bending along the zigzag direction as well as tearing along the same direction. Case (c) shows an example where bending occurs along an armchair configuration, observed in the cases where an upward force is applied.

**CONCLUSIONS**

We show how using a STM tip it is possible to create triangular flakes of graphene of tens of nanometers on top of graphite. From the analysis of our experiments and computer simulations we understand that the formation of these flakes usually involves a two-step process: first a voltage must be applied at the edge of a terrace, and then the tip must be scanned perpendicularly to the edge. The electrostatic force produced by the applied voltage is able to lift and tear the first graphite layer. This layer can then be easily lifted and displaced during the scanning producing the final flake observed in the STM. The correct understanding of these phenomena now allows us to have control on the electrostatic modification of a graphite or graphene surface and, moreover, shows a new route for the manipulation of any other layered materials using the same physical principles.

**Author contributions**

C. R-V, G. S-A and D. C. Milan performed the experimental work. C. U. supervised and performed part of the experimental work. J. M-A performed the molecular dynamics simulations. M. J. C. supervised and performed part of the molecular dynamics simulations. M. Moaied performed the ab initio calculations. J. J. Palacios supervised and performed part of the ab initio calculations. C. U. and M. J. C. wrote the first draft of the manuscript. All authors contributed to the writing of the final version of the manuscript.

**Acknowledgements**

We thank J. Fernández-Rossier and Albert Guijarro for their insightful comments. Simulations were performed in the computer cluster and other computer resources of the Dept. of Applied Physics at the UA. GSA acknowledges the support of the CONICIT. This work is supported by the Generalitat Valenciana through grants references PROMETEO2012/011 and FPA/2013/A/081 and the Spanish government MINECO through grants FIS2010-21883, FIS2013-47328, the Comunidad de Madrid Progrmas S2013/MIT-3007 and P2013/MIT-2850 and CONSOLIDER CSD2007-0010. The authors thankfully acknowledge the computer resources, technical expertise and assistance provided by the Centro de Computación Científica of the Universidad Autónoma de Madrid.